\documentclass[prd,twocolumn,floatfix,amsmath,nofootinbib,amssymb,floatfix]{revtex4}
\usepackage{graphicx,color,dcolumn,booktabs,bm}
\usepackage{longtable,lscape}
\usepackage{txfonts}
\usepackage{overpic}
\usepackage{amssymb}
\usepackage{indentfirst}
\usepackage{feynmf}   
\usepackage{slashed}  
\usepackage{cases}
\usepackage{color}
\usepackage{multirow}
\usepackage{slashed}
\usepackage{epstopdf}
\usepackage{longtable}
\usepackage{ulem}
\usepackage{graphicx,color,dcolumn,booktabs,bm}
\usepackage[colorlinks,
            citecolor=blue,
            anchorcolor=red,
            menucolor=red,
            linkcolor=red,
            filecolor=red,
            runcolor=red,
            urlcolor=blue,
            frenchlinks=red]{hyperref}

\graphicspath{{Figures/}} %
\allowdisplaybreaks

\begin{document}

\title{Broad resonance structure in $e^+e^-\to f_1(1285)\pi^+\pi^-$ and higher $\rho$-meson excitations}

\author{Xiang Liu$^{1,2,3,4}$}\email{xiangliu@lzu.edu.cn}
\author{Qin-Song Zhou$^{1,3}$}\email{zhouqs13@lzu.edu.cn}
\author{Li-Ming Wang$^{5,3}$}\email{lmwang@ysu.edu.cn}
\affiliation{$^1$School of Physical Science and Technology, Lanzhou University, Lanzhou 730000, China\\
$^2$Joint Research Center for Physics, Lanzhou University and Qinghai Normal University, Xining 810000, China\\
$^3$Lanzhou Center for Theoretical Physics, Key Laboratory of Theoretical Physics of Gansu Province,
and Frontiers Science Center for Rare Isotopes, Lanzhou University, Lanzhou 730000, China\\
$^4$Research Center for Hadron and CSR Physics, Lanzhou University $\&$ Institute of Modern Physics of CAS, Lanzhou 730000, China\\
$^5$Key Laboratory for Microstructural Material Physics of Hebei Province, School of Science,
Yanshan University, Qinhuangdao 066004, China}

\begin{abstract}
Recently, the BaBar Collaboration reported a broad resonance structure near 2 GeV when analyzing the $e^+e^-\to f_1(1285)\pi^+\pi^-$ process, which provides a good opportunity to study the higher $\rho$-mesonic states.
When considering the $\rho(1900)$ and $\rho(2150)$ contributions, the experimental data of the cross section of $e^+e^-\to f_1(1285)\pi^+\pi^-$ around 2 GeV can be well depicted, especially the observed broad resonance structure in $e^+e^-\to f_1(1285)\pi^+\pi^-$ containing two substructures is suggested by the present work. 
Additionally, we also indicate a possible signal of $\rho(5S)$ around 2.5 GeV in the $e^+e^-\to f_1(1285)\pi^+\pi^-$ process. The obtained result may provide a valuable hint to identify higher $\rho$ mesons, which will be a new task for the BESIII and Belle II experiments with the accumulation of higher precision data.
\end{abstract}
\date{\today}
\maketitle

\section{Introduction}\label{sec1}

\begin{figure}[htbp]
  \centering
  \includegraphics[width=240pt]{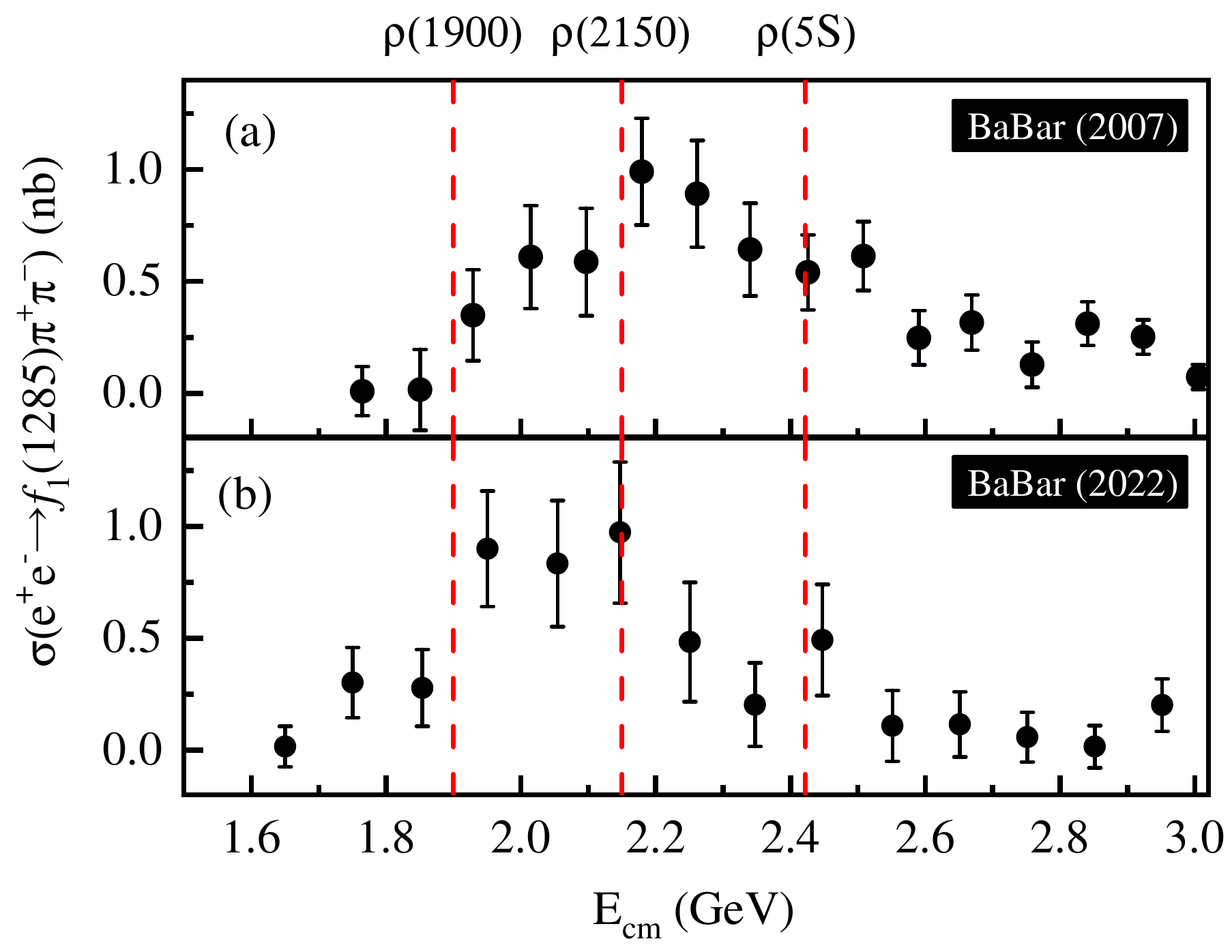}\\
  \caption{The experimental data of $e^+ e^- \to f_1(1285)\pi^+\pi^-$. Here, the experimental data shown in (a) were measured by the BaBar Collaboration in 2007 \cite{BaBar:2007qju}, and the result in (b) was newly measured by the BaBar Collaboration in 2022 \cite{BaBar:2022ahi}.}\label{ExpData}
\end{figure}

Very recently, the BaBar Collaboration announced the measurement of the $e^+e^-\to K^+K^-3\pi^0$, $e^+e^-\to K_S^0 K^\pm \pi^\mp 2\pi^0$, and $e^+e^-\to K_S^0K^\pm \pi^\mp \pi^+\pi^-$ processes
at center-of-mass energies from threshold to 4.5 GeV, where this analysis was performed with the initial state radiation method \cite{BaBar:2022ahi}. By the $e^+e^-\to K_S^0K^\pm \pi^\mp \pi^+\pi^-$ reaction, BaBar reconstructed the  $e^+e^-\to f_1(1285)\pi^+\pi^-$ process, by which a broad enhancement structure around 2 GeV was observed with resonance parameters 
\begin{eqnarray}\nonumber
\rm{M}=2.09\pm0.03\,{\rm GeV}, \quad
\Gamma=0.50\pm 0.06\,{\rm GeV}.\nonumber
\end{eqnarray}
Actually, in 2007, the BaBar Collaboration had already studied the $e^+e^-\to f_1(1285)\pi^+\pi^-$ process \cite{BaBar:2007qju}, and a broad resonance structure was observed around 2 GeV by a single Breit-Wigner fitting, which has resonance parameters
\begin{eqnarray}\nonumber
\rm{M}=2.15\pm0.04\pm0.05\,{\rm GeV},\quad
\Gamma=0.35\pm 0.04\pm0.05\,{\rm GeV}.\nonumber
\end{eqnarray}
Here, the experimental results for the cross section of $e^+e^-\to f_1(1285)\pi^+\pi^-$ are collected in Fig. \ref{ExpData}.
Because of the constraint of conservations of angular momentum and $P$ parity, it requires that the $f_1(1285)\pi^+\pi^-$ system must have a quantum number $I^G(J^{PC})=1^+(1^{--})$.
Therefore, the measured $e^+e^-\to f_1(1285)\pi^+\pi^-$ process may provide a good chance to construct the $\rho$-meson family.

\begin{figure*}[hptb]
  \centering
  \begin{tabular}{cc}
  \includegraphics[width=200pt]{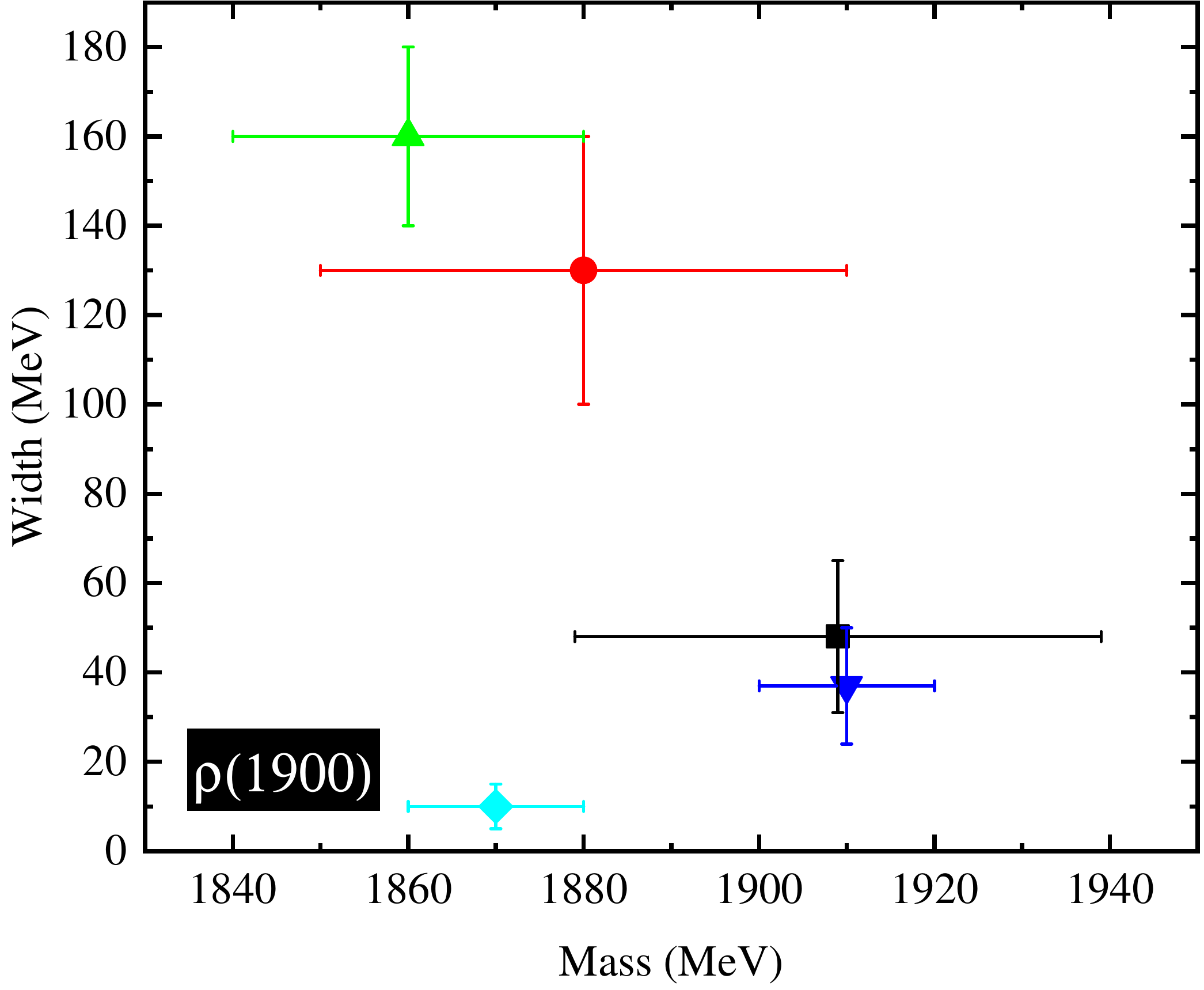}&\includegraphics[width=281pt]{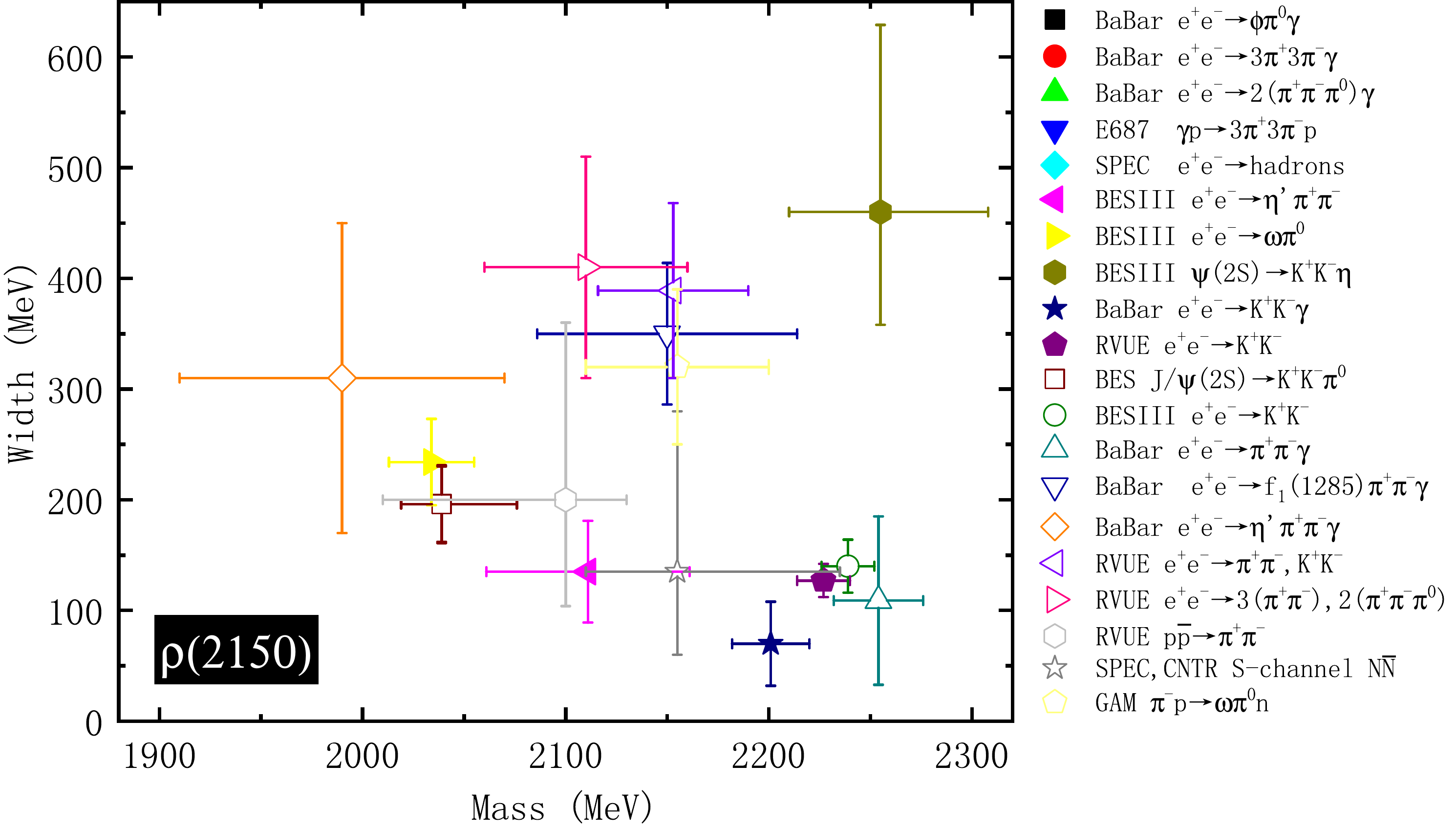}
  \end{tabular}
  \caption{The resonance parameters of the $\rho(1900)$ and $\rho(2150)$ are measured by different experiments collected in PDG \cite{ParticleDataGroup:2020ssz}.}\label{RP}
\end{figure*}

In the mass region around 2 GeV, there are three $\rho$-mesonic states collected in the Particle Data Group (PDG) \cite{ParticleDataGroup:2020ssz}, which are the $\rho(1900)$, $\rho(2150)$, and $\rho(2000)$.
Although experiments have reported some $\rho$ states around 2 GeV in the past few decades \cite{ParticleDataGroup:2020ssz}, categorizing these states into the $\rho$-meson family is far from being established.
In Fig. \ref{RP}, we list the resonance parameters of the $\rho(1900)$ and $\rho(2150)$ measured by different experiment groups \cite{ParticleDataGroup:2020ssz}, by which the messy situation of the measured resonance parameters of these two higher $\rho$ states around 2 GeV is presented. 
Evidently, the resonance parameters of the $\rho(1900)$ and $\rho(2150)$ from different experiments are different.
If considering the experimental uncertainties, the obtained widths of the $\rho(1900)$ and $\rho(2150)$ are in the ranges of $5-180$ MeV and $32-630$ MeV, respectively. 
The $\rho(2000)$ is collected in PDG as the “further state” \cite{ParticleDataGroup:2020ssz}, which was first reported by an amplitude analysis of the data of $p\bar{p}\to\pi\pi$ \cite{Hasan:1994he}. After that, its existence was confirmed by a combined analysis of the $e^+e^-\to\omega\eta\pi^0$ and $\omega\pi$ \cite{Bugg:2004xu}.
However, many recent experimental measurements with higher precision for the $\rho$-mesonic states show that there is an enhancement structure with the width around 100 MeV and the mass near 2 GeV rather than a broad structure with the width larger than 300 MeV \cite{BESIII:2020xmw,BESIII:2020kpr}.
On the theoretical side, various models have also been used to study the mass spectrum of $\rho$ mesons, including potential model \cite{Godfrey:1985xj,Barnes:1996ff,Ebert:2009ub,Li:2021qgz,Wang:2021gle,Wang:2021abg}, Regge trajectory \cite{He:2013ttg,Feng:2021igh}, and other methods \cite{Branz:2010ub,Hilger:2015ora,Yu:2021ggd}, where the $\rho(1900)$, $\rho(2150)$, and $\rho(2000)$ are usually categorized as $\rho(3^3S_1)\equiv\rho(3S)$, $\rho(4^3S_1)\equiv\rho(4S)$, and $\rho(2^3D_1)\equiv\rho(2D)$, respectively.
The quark pair creation (QPC) model is widely applied to estimate Okubo-Zweig-Iizuka (OZI)-allowed two-body strong decays of a hadron.
By using the QPC model \cite{Li:2021qgz,Wang:2021gle,Wang:2021abg,He:2013ttg,Feng:2021igh}, the widths of higher $\rho$ mesons were obtained, where the calculated widths of the $\rho(1900)$, $\rho(2105)$, and $\rho(2000)$ are in the ranges of $125-184$ MeV, $102-122$ MeV, and $179-228$ MeV, respectively.

As mentioned above, we briefly review the status of experimental and theoretical research on $\rho$-mesonic states around 2 GeV. When facing the BaBar's measurement of the cross section of the $e^+e^-\to f_1(1285)\pi^+\pi^-$ process, some questions are raised as shown below.
Usually, the theoretical results show that the widths of the $\rho(1900)$ and $\rho(2150)$ are about 100 MeV, and the width of the $\rho(2000)$ is less than 300 MeV. Obviously, the broad structure observed by the BaBar Collaboration in the $e^+e^-\to f_1(1285)\pi^+\pi^-$ process is not consistent with these reported $\rho$ states. 
Can we find a solution to alleviate such inconsistency?

To clarify this issue, we conjecture that the reported broad structure in the $f_1(1285)\pi^+\pi^-$ invariant mass spectrum can be still due to the interference effect from some higher $\rho$-mesonic states. According to the theoretical study of higher $\rho$-mesonic states around and above 2 GeV by the modified Godfrey-Isgur model \cite{Li:2021qgz,Wang:2021gle,Wang:2021abg}, 
we suggest that the $\rho(3S)$, $\rho(4S)$ and $\rho(5S)$ contribution should be introduced when depicting the data of the $e^+e^-\to f_1(1285)\pi^+\pi^-$ process, where the $\rho(3S)$ and $\rho(4S)$ correspond to the $\rho(1900)$ and $\rho(2150)$, respectively. 
In realistic calculation, we adopt the effective Lagrangian approach to calculate the cross section of $e^+e^-\to \rho^*\to f_1(1285)\pi^+\pi^-$. By 
fitting the experimental data of the cross section of $e^+e^-\to f_1(1285)\pi^+\pi^-$, we find that the reported broad enhancement structure in $e^+e^-\to f_1(1285)\pi^+\pi^-$ may contain two substructures, i.e., the $\rho(1900)$ and $\rho(2150)$. In addition, we also find  possible evidence of $\rho(5S)$ when reproducing the line shape of the cross section of $e^+e^-\to f_1(1285)\pi^+\pi^-$ well. With more precise data collected by BESIII and Belle II in the near future, we have a reason to believe that our observations can be tested.

This paper is organized as follows. After the introduction, we illustrate the study of the $e^+e^-\to  f_1(1285)\pi^+\pi^-$ process by considering the intermediate higher $\rho$ mesons as shown in Sec. \ref{sec2}. 
In Sec. \ref{sec3}, we present numerical results for the analysis of the cross section of the $e^+e^-\to  f_1(1285)\pi^+\pi^-$ process.
Finally, the paper ends with a discussion and conclusion in Sec. \ref{sec4}.

\section{Higher $\rho$-mesonic state contributions to $e^+e^-\to  f_1(1285)\pi^+\pi^-$}\label{sec2}

\begin{figure}[htbp]
  \centering
  \begin{tabular}{cc}
  \includegraphics[width=120pt]{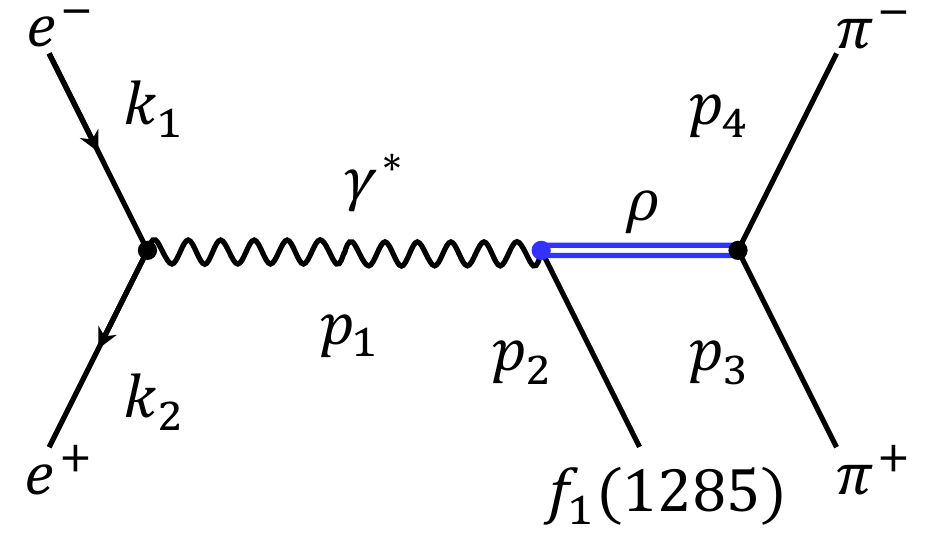}&\includegraphics[width=120pt]{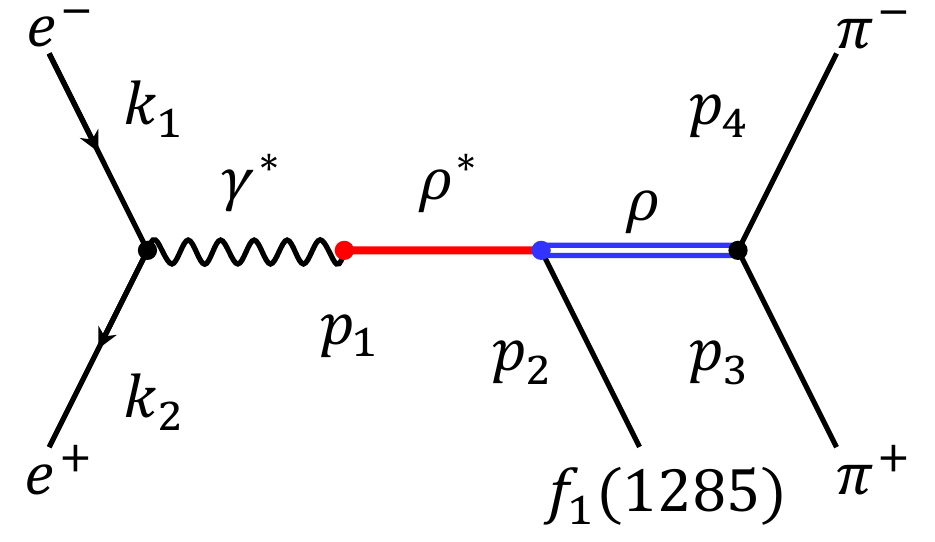}\\
  (a)&(b)\\
  \end{tabular}
  \caption{The schematic diagrams depict the reaction $e^+ e^- \to f_1(1285)\pi^+\pi^-$. Here, (a) depicts the virtual photon directly coupling to $f_1(1285) \rho$, while  (b) is due to the intermediate state $\rho^*$ contribution, where $\rho^*$ denotes higher $\rho$-mesonic states.}\label{Fey}
\end{figure}

Since the scatter plot of $m(\pi^+\pi^-)$ vs $m(K_{S}^{0}K^{\pm}\pi^{\pm})$ given by the BaBar Collaboration \cite{BaBar:2022ahi} shows that $\pi^+\pi^-$ is dominantly from $\rho(770)$, we can assume that the reaction $e^+e^-\to f_1(1285) \pi^+\pi^-$ occurs mainly through the two reaction mechanisms shown in Fig. \ref{Fey}.
Here, Fig. \ref{Fey} (a) depicts the virtual photon directly coupled with $f_1(1285) \rho$, which provides the background contribution,  while Fig. \ref{Fey} (b) reflects intermediate higher $\rho$-meson contribution to $e^+e^-\to f_1(1285) \pi^+\pi^-$.

In this work, we adopt the effective Lagrangian approach to describe these interaction vertices shown in Fig. \ref{Fey}. 
The effective Lagrangians involved in the concrete calculation include \cite{Rosenberg:1962pp,Kochelev:1999zf,Kaymakcalan:1983qq,Bauer:1975bv,Bauer:1975bw}

\begin{eqnarray}
\mathcal{L}_{\gamma f_1 \rho}&=&g_{\gamma f_1\rho}\epsilon^{\mu\nu\alpha\beta}(\partial_{\mu}A_{\alpha}\partial^{\lambda}\partial_{\lambda}\rho_{\nu}-\partial^{\lambda}\partial_{\lambda}A_{\alpha}\partial_{\mu}\rho_{\nu})f_{1\beta},\\
\mathcal{L}_{\rho^* f_1 \rho}&=&g_{\rho ^* f_1\rho}\epsilon^{\mu\nu\alpha\beta}(\partial_{\mu}\rho^{*}_{\alpha}\partial^{\lambda}\partial_{\lambda}\rho_{\nu}-\partial^{\lambda}\partial_{\lambda}\rho^{*}_{\alpha}\partial_{\mu}\rho_{\nu})f_{1\beta},\\
\mathcal{L}_{\rho\pi\pi}&=&g_{\rho\pi\pi}\rho^{\mu}\pi\overleftrightarrow{\partial_{\mu}}\pi,\\
\mathcal{L}_{\gamma \rho^*}&=&-e\frac{m_{\rho^*}^2}{f_{\rho^*}}\rho^{*\mu}A_{\mu},
\end{eqnarray}
where $\rho^*$, $\rho$, $f_1$, and $\pi$ stand for the mesonic fields of higher $\rho$ meson, $\rho(770)$, $f_1(1285)$, and $\pi$, respectively.
With the above preparation, the amplitude of $e^+ e^- \to f_1(1285)\pi^+\pi^-$ corresponding to Fig. \ref{Fey} (a) is written as
\begin{eqnarray} \nonumber
\mathcal{M}_{\rm{Dir}}&=&g_{\gamma^*f_1 \rho}g_{\rho\pi\pi}\bar{v}(k_2)(ie\gamma_\sigma)u(k_1)\frac{-g^{\sigma\alpha}}{p_1^2}\epsilon^{\mu\nu\alpha\beta}(p_{1\mu}q^2+p_1^2 q_{\mu})\\
&&\times\frac{\tilde{g}_{\nu\kappa}(q)}{q^2-m_{\rho}^2+im_{\rho}\Gamma_{\rho}}(p_{4}^{\kappa}-p_{3}^{\kappa})\varepsilon^{*\beta}(p_2)\mathcal{F}(s),
\end{eqnarray}
where $p_2$, $p_3$, and $p_4$ are the four momenta of final states $f_1(1285)$, $\pi^+$, and $\pi^-$, respectively. And, $p_1=k_1+k_2$, $q=p_3+p_4$, $\tilde{g}^{\nu\kappa}(q)=-g^{\nu\kappa}+q^{\nu}q^{\kappa}/q^2$, and the form factor $\mathcal{F}(s)=e^{-b[\sqrt{s}-(m_{f_1(1285)}+m_{\pi^+}+m_{\pi^-})]}$ are defined.
The amplitude of the reaction of $e^+ e^- \to f_1(1285)\pi^+\pi^-$ via the intermediate higher $\rho$ state, which corresponds to Fig. \ref{Fey} (b), can be written as the product of the amplitude of $e^+e^-\to\rho_i^*$ and the amplitude of $\rho_i^*\to f_1(1285)\pi^+\pi^-$:
\begin{eqnarray}
\mathcal{M}_{\rho_i^*}=\frac{\mathcal{M}_{e^+e^-\to\rho_i^*}
\mathcal{M}_{\rho_i^*\to f_{1}(1285)\pi^+\pi^-}}{p_1^2-m_{\rho_i^*}^2+im_{\rho_i^*}\Gamma_{\rho_i^*}},
\end{eqnarray}
where the amplitude of $e^+e^-\to \rho_i^*$ is written as
\begin{eqnarray}
\mathcal{M}_{e^+e^-\to\rho_i^*}&=&\bar{v}(k_2)(ie\gamma_\sigma)u(k_1)\frac{-g^{\sigma\alpha}}{p_1^2}\frac{-e m_{\rho_i^*}^2}{f_{\rho_i^*}}\varepsilon^{*\alpha}(p_1)
\end{eqnarray}
and the amplitude of $\rho_i^*\to f_1(1285)\pi^+\pi^-$ is expressed as 
\begin{eqnarray}\nonumber
\mathcal{M}_{\rho_i^*\to f_{1}(1285)\pi^+\pi^-}&=&g_{\rho_i^* f_1(1285)\rho}g_{\rho\pi\pi}\varepsilon^{\alpha}(p_1)\epsilon^{\mu\nu\alpha\beta}(p_{1\mu}q^2+p_1^2 q_{\mu})\\
&&\times\frac{\tilde{g}_{\nu\kappa}(q)}{q^2-m_{\rho}^2+im_{\rho}\Gamma_{\rho}}(p_{4}^{\kappa}-p_{3}^{\kappa})\varepsilon^{*\beta}(p_2).
\end{eqnarray}
With the partial decay width of $\rho(770) \to \pi\pi$ listed in PDG \cite{ParticleDataGroup:2020ssz}, we estimate the coupling constant $g_{\rho\pi\pi}=6$ $\rm{GeV}^{-2}$.

The differential cross section of $e^+e^-\to f_1(1285)\pi^+\pi^-$ is
\begin{eqnarray} 
d\sigma=\frac{1}{32(2\pi)^5 \sqrt{s}\sqrt{k_1 k_2}}|\overline{\mathcal{M}_{\rm{Total}}|^2}|\vec{p}_2||\vec{p}_3^*|d\Omega_2 d\Omega_3^* dm_{34}, \label{cross section}
\end{eqnarray} 
where $\vec{p_2}$ ($\Omega_2$) is the three-momentum (solid angle) of $f_1(1285)$ in the center-of-mass frame, $\vec{p}_3^*$ ($\Omega_3^*$) stands for the three-momentum (solid angle) of the $\pi^+$ in the rest frame of $\pi^+$ and $\pi^-$, and $m_{34}$ is the invariant mass of the
$\pi^+$ and $\pi^-$ system. On the other hand, the total amplitude of $\mathcal{M}_{\rm{Total}}$ is
\begin{eqnarray}
\mathcal{M}_{\rm{Total}}=\mathcal{M}_{\rm{Dir}}+\sum_{\rho_i^*}\mathcal{M}_{\rho_i^*}e^{i\phi_{\rho_i^*}}\label{ampT}.
\end{eqnarray}
Here, $\rho_i^*$ denote the allowed higher $\rho$-meson states, while $\phi_{\rho_i^*}$ is the phase angle among different amplitudes, which is determined by fitting experimental data.
In addition, these coupling constants included in the amplitudes are also determined by fitting experimental data.

To facilitate comparison with the experimental results, the combined branching ratio $\Gamma_{e^+e^-}\mathcal{B}(\rho_{i}^*\to f_1(1285)\pi^+\pi^-)$ can be calculated by using Eqs. (\ref{dilipton width}) and (\ref{decay width}) when the coupling constants are obtained by fitting the experimental data, where $\Gamma_{e^+e^-}$ is the dilepton decay width of $\rho_i^{*}$, i.e., 
\begin{eqnarray}
\Gamma_{e^+e^-}=\frac{e^4 m_{\rho_{i}^*}}{12\pi f_{\rho_{i}^{*}}^2}.\label{dilipton width}
\end{eqnarray}
Then, the $\mathcal{B}(\rho_{i}^*\to f_1(1285)\pi^+\pi^-)=\Gamma(\rho_{i}^*\to f_1(1285)\pi^+\pi^-)/\Gamma_{\rho_i^*}$ denotes the branching ratio of $\rho_{i}^*\to f_1(1285)\pi^+\pi^-$
with the expression 
\begin{eqnarray}\nonumber
\Gamma(\rho_{i}^*\to f_1(1285)\pi^+\pi^-)&=&\frac{1}{(2\pi)^5}\frac{1}{16m_{\rho_i^*}^2}\overline{|\mathcal{M}_{\rho_i^*\to f_{1}(1285)\pi^+\pi^-}|^2}\\ 
&&\times |\vec{p}_2||\vec{p}_3^*|d\Omega_2 d\Omega_3^* dm_{34}.\label{decay width}
\end{eqnarray}
Here, all symbols have the same meaning as in Eq. (\ref{cross section}).

\section{Numerical results}\label{sec3}

In this section, we adopt two schemes to fit the experimental data of the cross section of $e^+e^-\to f_1(1285)\pi^+\pi^-$ measured by the BaBar Collaboration \cite{BaBar:2022ahi,BaBar:2007qju}.
In scheme 1, we consider the contributions of these higher $\rho$-mesonic states near 2 GeV, which include the $\rho(1900)$, $\rho(2150)$, and $\rho(2000)$, to decipher the broad resonance structure existing in the $e^+e^-\to f_1(1285)\pi^+\pi^-$ process. We also notice that 
there is an event accumulation around 2.5 GeV. Thus, in scheme 2, $\rho(5S)$ associated with the $\rho(1900)$, $\rho(2150)$, and $\rho(2000)$
is considered for depicting the data well. 
In our investigation, these coupling constants, relative phase angles, the resonance parameters of relevant higher $\rho$-mesonic states, and the parameter $b$ in form factor are treated as the fitting parameters.

\begin{figure}[hptb]
  \centering
  \includegraphics[width=240pt]{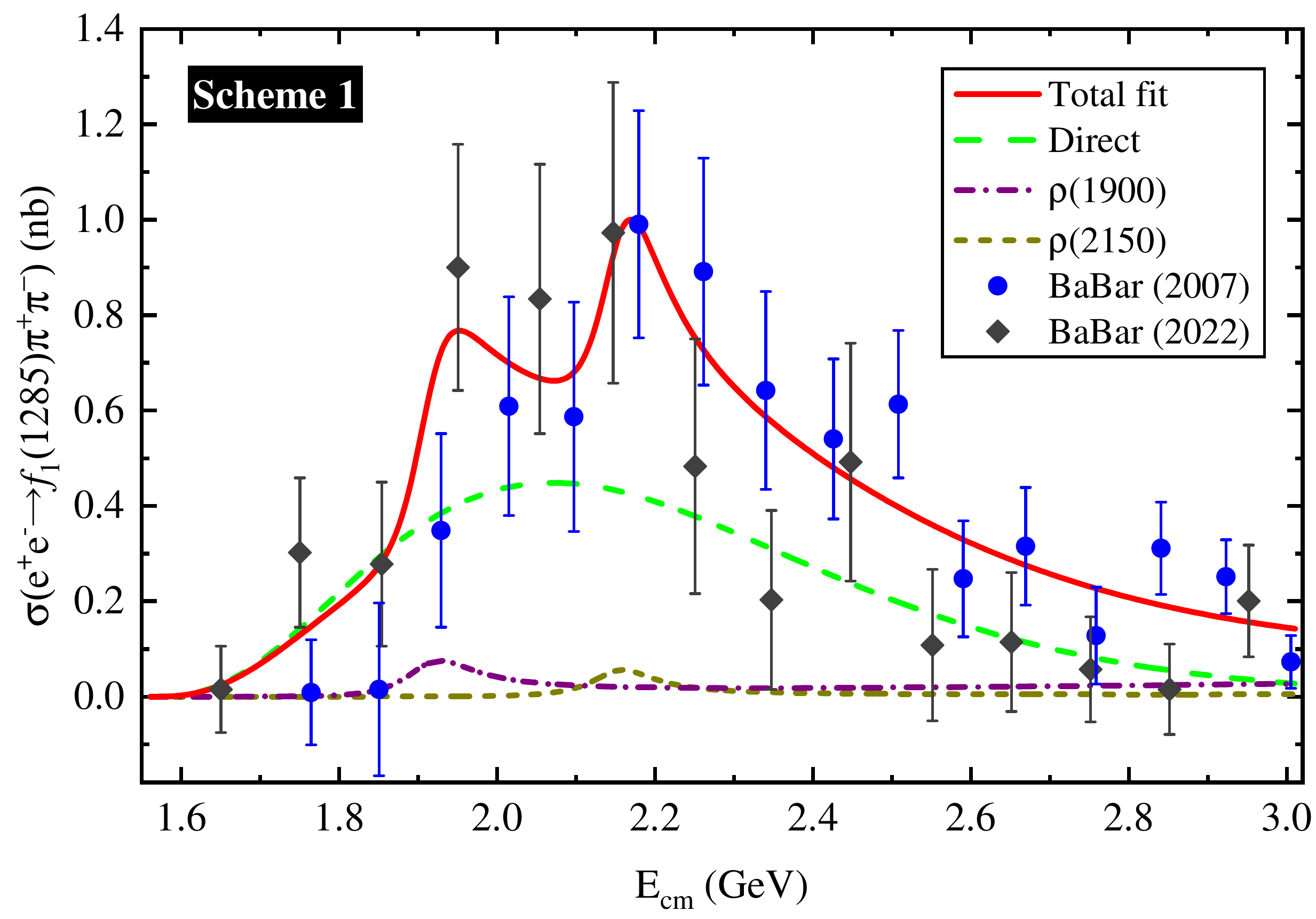}
  \caption{The fit to the experimental data of the cross sections of $e^+e^-\to f_1(1285)\pi^+\pi^-$ measured by the BaBar Collaboration in 2007  \cite{BaBar:2007qju} (blue dots with error bars) and in 2022 \cite{BaBar:2022ahi} (black dots with error bars). In this scheme, only the contributions from the $\rho(1900)$ and $\rho(2150)$ are considered. Here, the calculated value of $\chi^2/\rm{n.d.f.}$ is 1.63.}\label{FitResult2RF}
\end{figure}

Although there are three reported $\rho$ states near 2 GeV, the $\rho(2000)$ as a typical D-wave state
\cite{Li:2021qgz,Wang:2021gle,He:2013ttg,Yu:2021ggd} is not included in our study. The main reason is that the dilepton width of the D-wave vector meson is suppressed compared to the case of the S-wave vector meson \cite{Godfrey:1985xj,Wang:2021gle,Wang:2020kte,Zhou:2022ark}.
Thus, in scheme 1, we consider only the contributions from the $\rho(1900)$ and $\rho(2150)$ to reproduce the cross section data of $e^+e^-\to f_1(1285)\pi^+\pi^-$ process.
The fitted result of the cross sections of $e^+e^-\to f_1(1285)\pi^+\pi^-$ is shown in Fig. \ref{FitResult2RF}, where these obtained fitting parameters are collected into Table \ref{FitResult2RT}. For convenience,  $g_{\rho_{i}^{*}f_{1}(1285)\rho}g_{\rho\pi\pi}/f_{\rho_{i}^*}$ is abbreviated to be  $g_{\rho_i^*}$. 
With the central values of the fitted parameters, the value of $\chi^2/\rm{n.d.f.}$ is obtained to be 1.63. 

As shown in Fig. \ref{FitResult2RF}, the fitted curve can well reproduce the experimental data of the cross sections for $e^+e^-\to f_1(1285) \pi^+\pi^-$. Especially, our study shows that the reported broad enhancement structure in $e^+e^-\to f_1(1285)\pi^+\pi^-$ should  at least contain two substructures like the $\rho(1900)$ and $\rho(2150)$. We also list 
the fitted results of the resonance parameters for both $\rho(1900)$ and $\rho(2150)$ in Table \ref{FitResult2RT}, which are within the range of previous theoretical results \cite{Ebert:2009ub,Li:2021qgz,Wang:2021gle,He:2013ttg,Feng:2021igh}. 

By such an effort, we provide a possible explanation to understand why there exists the broad structure with width around 300 MeV in $e^+e^-\to f_1(1285)\pi^+\pi^-$, which is caused by the interference effect of the $\rho(1900)$ and $\rho(2150)$. Thus, the inconsistency 
of the BaBar observation of very broad enhancement structure \cite{BaBar:2022ahi,BaBar:2007qju} and the established
$\rho$ states near 2 GeV in PDG can be naturally alleviated.

\begin{table}[htbp]
  \centering
  \caption{The parameters involved in scheme 1, which are obtained by fitting the cross sections of the $e^+e^-\to f_1(1285) \pi^+\pi^-$ process. Here, the experimental data are from BaBar \cite{BaBar:2022ahi,BaBar:2007qju}.}\label{FitResult2RT}
  \begin{tabular}{ccc}
  \toprule[1pt]
  \midrule[1pt]
  Parameters & $\quad$ & Values   \\
  \midrule[1pt]
  $m_{\rho(1900)}$ (MeV) &  $\quad$ &$1913\pm34$\\
  $m_{\rho(2150)}$ (MeV) &  $\quad$ &  $2151\pm32$\\
  $\Gamma_{\rho(1900)}$ (MeV) &  $\quad$ &$106\pm21$\\
  $\Gamma_{\rho(2150)}$ (MeV) &  $\quad$ &  $98\pm16$\\
  $g_{\gamma^*\rho f_1(1285)}$ $(\rm{GeV}^{-2})$& $\quad$ & $4.11\pm 0.38$\\
  $g_{\rho(1900)}$ $(\rm{GeV}^{-2})$ & $\quad$ & $0.088\pm0.009$\\
  $g_{\rho(2150)}$  $(\rm{GeV}^{-2})$ & $\quad$ & $0.024\pm0.007$\\
  $b$ $(\rm{GeV}^{-1})$ & $\quad$ & $3.74\pm0.18$\\
  $\phi_{\rho(1900)}$ (rad)& $\quad$ & $0.63\pm 0.02$\\
  $\phi_{\rho(2150)}$ (rad)& $\quad$ & $0.90\pm0.04$\\
  $\chi^2/\rm{n.d.f.}$ & $\quad$ & 1.63\\
\midrule[1pt]
\bottomrule[1pt]
\end{tabular}
\end{table}

\begin{figure}[hptb]
  \centering
  \includegraphics[width=240pt]{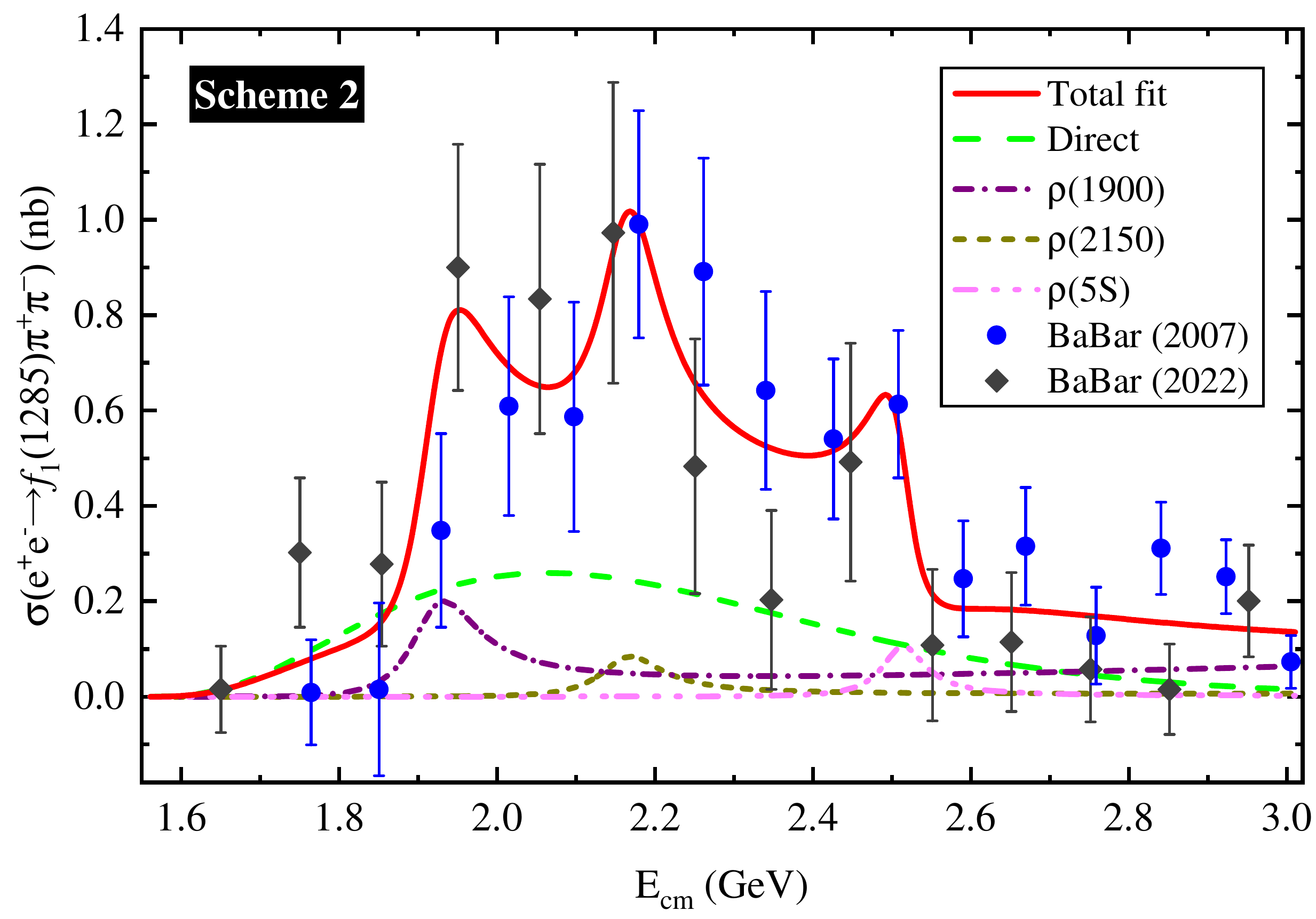}
  \caption{The fit to the experimental data of the cross sections of $e^+e^-\to f_1(1285)\pi^+\pi^-$ measured by the BaBar Collaboration in 2007  \cite{BaBar:2007qju} (blue dots with error bars) and 2022 \cite{BaBar:2022ahi} (black dots with error bars). In this scheme, the contributions from $\rho(1900)$, $\rho(2150)$, and $\rho(5S)$ are considered. Here, the calculated value of $\chi^2/\rm{n.d.f}$ is 1.52.}\label{FitResult3RF}
\end{figure}

Besides explaining the broad structure reported by BaBar, we notice an event cluster around 2.5 GeV in the measured cross section data. Thus, we propose to include the predicted $\rho(5S)$ state \cite{Li:2021qgz,Wang:2021gle,Wang:2021abg} associated with the $\rho(1900)$ and $\rho(2150)$, which is as scheme 2 in the present work. 
The fitted results are shown in Fig. \ref{FitResult3RF}, and the corresponding parameters are listed in Table \ref{FitResult3RT}. Obviously, the fitting result to the cross section of 
$e^+e^-\to f_1(1285)\pi^+\pi^-$ can be improved in scheme 2. We obtain the resonance parameter of the $\rho(5S)$ state, i.e., 
\begin{eqnarray}\nonumber
\rm{M}=2512\pm56\,{\rm MeV}, \quad
\Gamma=66\pm 38\,{\rm MeV},\nonumber
\end{eqnarray}
which is consistent with theoretical predictions in Refs. \cite{Li:2021qgz,Wang:2021gle,Wang:2021abg}. 
We suggest future experiment to identify 
the $\rho(5S)$ signal with more precise data, which is crucial to construct $\rho$ meson family.


\begin{table}[htb]
  \centering
  \caption{The parameters for scheme 2 when fitting the cross sections of the $e^+e^-\to f_1(1285) \pi^+\pi^-$ process \cite{BaBar:2022ahi,BaBar:2007qju}.}\label{FitResult3RT}
  \begin{tabular}{ccc}
  \toprule[1pt]
  \midrule[1pt]
  Parameters & $\quad$ & Values   \\
  \midrule[1pt]
  $m_{\rho(1900)}$ (MeV) &  $\quad$ &$1919\pm38$\\
  $m_{\rho(2150)}$ (MeV) &  $\quad$ &  $2162\pm 31$\\
   $m_{\rho(5S)}$ (MeV) &  $\quad$ &  $2512\pm56$\\
  $\Gamma_{\rho(1900)}$ (MeV) &  $\quad$ &$102\pm41$\\
  $\Gamma_{\rho(2150)}$ (MeV) &  $\quad$ &  $95\pm28$\\
  $\Gamma_{\rho(5S)}$ (MeV) &  $\quad$ &  $66\pm38$\\
  $g_{\gamma^*\rho f_1(1285)}$ $(\rm{GeV}^{-2})$& $\quad$ & $3.19\pm 0.10$\\
  $g_{\rho(1900)}$ $(\rm{GeV}^{-2})$ & $\quad$ & $0.133\pm0.019$\\
  $g_{\rho(2150)}$  $(\rm{GeV}^{-2})$ & $\quad$ & $0.028\pm0.008$\\
  $g_{\rho(5S)}$  $(\rm{GeV}^{-2})$ & $\quad$ & $0.007\pm0.002$\\
  $b$ $(\rm{GeV}^{-1})$ & $\quad$ & $3.78\pm0.49$\\
  $\phi_{\rho(1900)}$ (rad)& $\quad$ & $0.52\pm 0.21$\\
  $\phi_{\rho(2150)}$ (rad)& $\quad$ & $1.47\pm0.19$\\
  $\phi_{\rho(5S)}$ (rad)& $\quad$ & $3.20\pm 0.36$\\
  $\chi^2/\rm{n.d.f}$ & $\quad$ & 1.52\\
\midrule[1pt]
\bottomrule[1pt]
\end{tabular}
\end{table}

To facilitate comparison with the experimental results, by substituting the values of the coupling constants listed in Table \ref{FitResult3RT} into Eqs. (\ref{dilipton width}) and (\ref{decay width}), we can estimate the values of the combined branching ratio $\Gamma_{e^+e^-}\mathcal{B}(\rho^*_{i}\to f_{1}(1285)\pi^+\pi^-)$, which are presented in Table \ref{CBR}. 
The values of these combined branching ratios can be measured by experiments with higher precision in the future, by which our scenario proposed in this work can be tested.

\begin{table}[htb]
  \centering
  \caption{The combined branching ratios $\Gamma_{e^+e^-}\mathcal{B}(\rho_{i}^*\to f_1(1285)\pi^+\pi^-)$ are estimated by substituting the values of the coupling constants listed in Table \ref{FitResult3RT} into Eqs. (\ref{dilipton width}) and (\ref{decay width}).}\label{CBR}
  \begin{tabular}{ccc}
  \toprule[1pt]
  \midrule[1pt]
  Parameters & $\quad$ & Values (eV)  \\
  \midrule[1pt]
  $\Gamma_{e^+e^-}\mathcal{B}(\rho(1900)\to f_{1}(1285)\pi^+\pi^-)$ & $\quad$ & $2.08- 3.69$ \\
  $\Gamma_{e^+e^-}\mathcal{B}(\rho(2150)\to f_{1}(1285)\pi^+\pi^-)$ & $\quad$ & $5.71-18.48$ \\
$\Gamma_{e^+e^-}\mathcal{B}(\rho(5^3S_1)\to f_{1}(1285)\pi^+\pi^-)$ &  $\quad$ & $0.46-13.04$\\
\midrule[1pt]
\bottomrule[1pt]
\end{tabular}
\end{table}

\section{Discussion and conclusion}\label{sec4}

Very recently, the BaBar Collaboration measured 
$e^+e^-\to f_1(128) \pi^+\pi^-$ again \cite{BaBar:2022ahi} and reported a very broad enhancement structure around 2 GeV in the $f_1(128) \pi^+\pi^-$
invariant mass spectrum. In 2007, BaBar once indicated a similar phenomenon \cite{BaBar:2007qju}.
If making a comparison of BaBar's observation with these reported $\rho$ states around 2 GeV collected by PDG \cite{ParticleDataGroup:2020ssz}, the broad enhancement structure from BaBar cannot 
correspond to these reported $\rho$ states like $\rho(1900)$, $\rho(2000)$, and $\rho(2150)$. This puzzling phenomenon stimulates our interest. 

Supported by the $\rho$-meson spectroscopy \cite{Li:2021qgz,Wang:2021gle,He:2013ttg,Yu:2021ggd}, we propose an interference picture of the $\rho(1900)$ and $\rho(2150)$ to depict the observed broad enhancement structure around 2 GeV in the $e^+e^-\to f_1(128) \pi^+\pi^-$. By this approach, the data of the cross section of $e^+e^-\to f_1(128) \pi^+\pi^-$ can be reproduced well. Naturally, we provide a natural solution to understand the puzzling phenomenon mentioned above. 

In fact, an event cluster around 2.5 GeV can be found if checking the experimental data of the cross section of $e^+e^-\to f_1(128) \pi^+\pi^-$. We notice that different theoretical groups predicted 
the existence of the $\rho(5S)$ state. Thus, we include the contribution of the $\rho(5S)$ state to fit the experimental data again, where the resonance parameter of the $\rho(5S)$ state can be obtained, which is consistent with the theoretical predictions \cite{Li:2021qgz,Wang:2021gle,Wang:2021abg}. Confirming this evidence is a potential issue for our experimental colleagues.

In summary, $e^+e^-\to f_1(128) \pi^+\pi^-$ is an ideal process to identify higher $\rho$-mesonic states. At present, the $\rho$-meson family is far from being established. As one part of the whole hadronic zoo, we have enough interest in finding out new higher $\rho$-meson states. With the running of Belle II and BESIII, experimentalists do not stop  observing new $\rho$ states. We hope that the present work may inspire some insights to construct light flavor vector meson family.

\vfil

\section*{Acknowledgments}
This work is supported by the projects funded by Science and Technology Department of Qinghai Province (No. 2020-ZJ-728), the China National Funds for Distinguished Young Scientists under Grant No. 11825503, National Key Research and Development Program of China under Contract No. 2020YFA0406400, the 111 Project under Grant No. B20063, and the National Natural Science Foundation of China under Grant No. 12047501.

\end{document}